\documentclass[prd,amssymb,amsmath]{revtex4}
\usepackage{epsfig}
\newcommand{\beq}{\begin{equation}}
\newcommand{\eeq}{\end{equation}}
\newcommand{\bqa}{\begin{eqnarray}}
\newcommand{\eqa}{\end{eqnarray}}
\newcommand{\fr}{\frac}

\begin{document}
\title{Comment on gr-qc/0309003: ``Comment on `Absence of trapped surfaces and singularities in cylindrical collapse' ''}
\author{S\'{e}rgio M. C. V. Gon\c{c}alves}
\affiliation{Department of Physics, Yale University, New Haven, Connecticut 06511}
\date{\today}
\begin{abstract}
Very recently, a ``Comment'' by Wang [gr-qc/0309003] on a paper by Gon\c{c}alves [Phys. Rev. D {\bf 65}, 084045 (2002)] appeared, claiming that Gon\c{c}alves' analysis of trapped surfaces in certain kinds of cylindrical spacetimes was incomplete. Specifically, Wang claims to have found a coordinate extension of the spacetime used by Gon\c{c}alves (the Einstein-Rosen spacetime) which contains trapped surfaces; in addition, Wang also claims that some such trapped surfaces are apparent horizons. Here, I comment on Wang's ``Comment'', and argue that, while Wang's spacetime extension appears to exist and contain trapped surfaces, it does not render Gon\c{c}alves' results incomplete in the sense Wang claims. I also show that, contrary to Wang's claim, his spacetime extension does {\em not} contain apparent horizons, i.e., it does not contain outer marginally trapped surfaces which are the outer boundary of a trapped region. Further peripheral comments by Wang are also commented on.
\end{abstract}
\maketitle

In order to address each of Wang's claims, I summarize (and quote) them below:

\begin{itemize}
\item {\bf Claim 1.} Gon\c{c}alves used three different criteria to test for trapped surfaces in the Einstein-Rosen spacetime; ``...his first criterion is incorrect, while his second and third are the same as Penrose's.'' \\
\item {\bf Claim 2.} ``...his analysis of the non-existence of trapped surfaces in vacuum is incomplete.'' \\
\item {\bf Claim 3.} ``...we present an example that is a solution to the vacuum Einstein field equations and satisfies all the regular conditions imposed by him. After extending the solution to the whole spacetime, both trapped surfaces and apparent horizons are found.''
\end{itemize}

In what follows, to avoid unnecessary repetition of equations, we refer the reader to Wang's ``Comment''~\cite{wang03} and Gon\c{c}alves' paper~\cite{goncalves02} for details of the model in question. The authors' names, Wang and Gon\c{c}alves, are henceforth abbreviated as ``W'' and ``G'', respectively. Unless otherwise noted, equation numbers refer to those in \cite{wang03}.

\subsection*{Comment on Claim 1.}

G used three different criteria in~\cite{goncalves02} to test for trapped surfaces in the Einstein-Rosen spacetime containing a cylindrical thin matter shell. G's first criterion considers surfaces defined by {\em proper circumference radius}, which is a {\em physical} quantity, measurable by timelike observers exterior to (or comoving with) the shell. The second criterion uses a definition by Hayward~\cite{hayward00}, the so-called {\em specific area radius}, which relies on the area radius of a unit Killing-length cylinder; this is a good geometric definition, but is operationally lacking---twofold---from a physical standpoint: (i) the specific area radius is just a {\em coordinate} radius, which changes under the rescaling of the Killing coordinate $z\rightarrow\alpha z$ as $r\rightarrow\alpha^{-1}r$, and (ii) external timelike observers can only measure proper circumferences (as used in the first criterion), but not coordinate radii. Finally, the third criterion uses outgoing `radial' null geodesics directly, instead of cylinders of symmetry (with respect to which the null geodesics are orthogonal). The three different criteria used by G in~\cite{goncalves02} are simply alternative, {\em but entirely equivalent} ways of checking for trapped surfaces: if one shows that trapped surfaces are absent/present then so must all others. W's comment ``...while his second and third are the same as Penrose's.'' is therefore trivial, since the criteria are the same by construction: they all check the divergence of future-oriented outgoing radial null geodesics in the spacetime; the third criterion does so directly, whereas the first two do so indirectly by looking at the 4-gradient of cylinders of symmetry.

Regarding the first part of Claim 1, ``...his first criterion is incorrect...'', W is correct in pointing out that Eq. (6) is wrong, but his explanation is, however, wrong. In Eqs. (7)--(13), W derives the expansions for future-oriented outgoing/ingoing null geodesics orthogonal to infinite cylinders with axis collinear with the translational Killing direction $z$, and then shows that this is equivalent to Hayward's specific area radius criterion. Equation (6), however, is {\em not} valid for the specific area radius criterion: it is valid only for the proper area radius criterion. In the former, one takes a cylinder with unit coordinate length, thus with proper length $\ell=e^{-\psi}$, whereas in the latter the cylinder has unit proper length, $\ell=1$. The two radii are different and thus their 4-gradient is, in general, different too. The reasons given after Eq. (15) for the incorrectness of Eq. (6) are therefore not valid. Equation (6), however, is incorrect; the correct expression is:
\beq
\theta_{\pm}=F(t,r)\left(\partial_{t}\pm e^{\psi-\gamma}\partial_{r}\right){\mathcal R},
\eeq
where $F(t,r)=dt/d\lambda>0$, and $\lambda$ is an affine parameter along future-pointing, outgoing, radial null geodesics with tangent vector field
\beq
u^{\mu}=\fr{dx^{\mu}}{d\lambda}=\left(F,Fe^{\psi-\gamma},0,0\right),
\eeq
in the standard Einstein-Rosen coordinates $\{t,r,z,\phi\}$.

\subsection*{Comment on Claim 2.}

In \cite{goncalves02}, G considers the spacetime manifold of the form 
\beq
M_{-}\cup \Sigma \cup M_{+},
\eeq
where $\Sigma$ is a (timelike) thin matter shell with cylindrical metric
\beq
ds^{2}_{\Sigma}=-d\tau^{2}+e^{2\psi_{\Sigma}(\tau)}dz^{2}+R^{2}(\tau)e^{-2\psi_{\Sigma}(\tau)}d\phi^{2},
\eeq
where $\tau$ is the proper time of an observer comoving with the shell, and $M_{\pm}$ are vacuum regions {\em globally} defined by the Einstein-Rosen metric:
\beq
ds^{2}=e^{2(\gamma-\psi)}(-dt^{2}+dr^{2})+e^{2\psi}dz^{2}+r^{2}e^{-2\psi}d\phi^{2},
\eeq
where $r\in{\mathbb R}_{0}^{+}$, $t,z\in{\mathbb R}$, $\phi\in[0,2\pi)$, and $\psi$ and $\gamma$ are functions of $t,r$ alone (note that there is a different set of coordinates $\{t_{\pm},r_{\pm}\}$ in each of the coordinate patches $M_{\pm}$; the Killing coordinates $\{z,\phi\}$ are trivially identified in both patches). That is, G considered spacetimes which are everywhere, but for the thin matter shell, described by the Einstein-Rosen metric.

One of G's results in \cite{goncalves02} is that there are no trapped surfaces in {\em any} portion of {\em the spacetime described above}. W writes ``...his analysis of non-existence of trapped surfaces in vacuum is incomplete.'' G's trapped surface analysis in \cite{goncalves02} was confined to spacetimes which are {\em globally} described by the Einstein-Rosen metric. {\em No claim about trapped surfaces was made in \cite{goncalves02} regarding more general spacetimes, i.e., spacetimes which are {\em not} globally covered by the Einstein-Rosen coordinates.} W appears to have found (I did not check all of his calculations, but the result is plausible) an example of a cylindrical~\footnote{As defined by: (i) two commuting spacelike Killing vector fields, one translational ($\partial_{z}$) with infinite open orbits, and the other azimuthal ($\partial_{\phi}$) closed $S^{1}$ orbits, and (ii) elementary flatness at the symmetry axis.} vacuum spacetime  which fails to be globally covered by the Einstein-Rosen coordinates. Although very contrived, W's example appears to provide an analytical extension of a particular case of the Einstein-Rosen spacetime. This simply means that the results of G in \cite{goncalves02} will not necessarily hold, since they only apply to spacetimes everywhere described by the Einstein-Rosen metric.

\subsection*{Comment on Claim 3.}

W writes ``...we present an example that is a solution to the vacuum Einstein field equations and satisfies all the regular conditions imposed by him. After extending the solution to the whole spacetime, both trapped surfaces and apparent horizons are found.'' Assuming W's calculations are correct, his example does contain trapped surfaces, but it does {\em not} contain apparent horizons. The standard definition of apparent horizon~\cite{wald} is a closed spacelike two-surface with the property that (i) it is the {\em outer} boundary of a total trapped region, and (ii) the divergence of future-oriented outgoing null geodesics orthogonal to the surface vanishes thereon (i.e., the surface is outer marginally trapped). In the standard asymptotically flat case, such surfaces have $S^{2}$ topology. However, the translational symmetry (of cylindrical spacetimes, among others) precludes apparent horizons from being homeomorphic to $S^{2}$, since one can always continuously deform any such surface along the symmetry direction---e.g., by cutting the surface along a two-plane orthogonal to the Killing direction and gluing the two parts by a topological cylinder of arbitrary length---whereby property (i) above is violated (i.e., there is no such thing as outer boundary along the symmetry direction). A natural modification of the definition of apparent horizon for ${\mathbb R}$-symmetric spacetimes is:

\vspace{0.2cm}

Definition: {\em Let $(M,{\mathbf g})$ be a four-dimensional Lorentzian spacetime admitting a globally defined spacelike Killing vector field of translational type (i.e., with infinite open orbits generated by the group $G_{1}={\mathbb R}$), ${\pmb \xi}$. An apparent horizon in $(M,{\mathbf g})$ is a topological $S^{1}\times{\mathbb R}$ spacelike two-surface which is outer marginally trapped, and is the outer boundary of a (non-compact) trapped region along the spacelike two-sector of the quotient Lorentzian spacetime induced by the orbits of the Killing vector field.}

\vspace{0.2cm}

For the particular case of cylindrical symmetry, the apparent horizon is also a geometric cylinder.

The example provided by W contains an outer marginally trapped surface ($\theta_{+}=0$), which is the {\em inner} boundary of a trapped region that extends all the way out to ``spacelike infinity'' [the quotation marks denote the fact that the asymptotics are not known in W's example; he provides only a schematic conformal diagram; this, however, is not detrimental in terms of defining---at least locally---notions of ``inner/outer'' and ``future/past'', since the $r$ and $t$ coordinates are well-defined in the regions of interest (cf. Figs 1 and 2 in \cite{wang03})] on any given Cauchy hypersurface. W calls such outer marginally trapped surface a ``future apparent horizon'', borrowing Hayward's terminology~\cite{hayward00}. As explained above, this is not the standard definition; using it very easily allows for numerous counter-examples of, e.g., the hoop conjecture and Ida's no-horizon theorem in $(2+1)$ gravity~\cite{ida00}. More than just a definitional issue, the precise meaning of {\em apparent horizon} is crucial for the formulation and veracity of numerous results in general relativity. Because of this, I detail below an example which contains an outer marginally trapped surface analogous to that of W's example, which is {\em not} an apparent horizon. For added clarity, the complete Penrose diagram is constructed by means of an explicit conformal transformation. 

\subsection*{Outer marginally trapped surfaces are not necessarily apparent horizons}

Consider the flat FRW metric in $(2+1)$ dimensions:
\beq
ds^{2}=-dt^{2}+a^{2}(t)(dr^{2}+r^{2}d\phi^{2}),
\eeq
where $t\in{\mathbb R}$, $r\in{\mathbb R}_{0}^{+}$, and $\phi\in[0,2\pi)$. The Einstein tensor is (where the dot denotes $\partial_{t}$):
\beq
G_{\mu\nu}=\mbox{diag}\left(\fr{\dot{a}^{2}}{a^{2}},-a\ddot{a},-r^{2}a\ddot{a}\right).
\eeq
Now, take a perfect fluid for the matter content, and assume the equation of state $p=\rho$:
\beq
T_{\mu\nu}=\rho u_{\mu}u_{\nu}+p(u_{\mu}u_{\nu}+g_{\mu\nu})=\rho(2u_{\mu}u_{\nu}+g_{\mu\nu}), \label{stress}
\eeq
where $u^{\mu}=\delta^{\mu}_{t}$ is the four-velocity of an observer comoving with the fluid. Einstein's equations are then
\bqa
\fr{\dot{a}^{2}}{a^{2}}&=&\rho, \\
-a\ddot{a}&=&a^{2}\rho.
\eqa
This is trivially solved to give
\beq
a^{2}(t)=c_{1}t+c_{2},\label{a}
\eeq
where $c_{1}\in{\mathbb R}, c_{2}\in{\mathbb R}_{0}^{+}$ are constants, fixed by the initial data: $a(0)=\sqrt{c_{2}}$, $\dot{a}(0)=c_{1}/(2\sqrt{c_{2}})$.

\subsubsection*{Trapped surfaces}

To examine the possible existence of trapped circles, let us consider future-oriented, outgoing, radial null geodesics, which are generated by the vector field
\beq
k^{\mu}=(a(t)F(t,r),F(t,r),0),
\eeq
for some positive-definite real-valued function $F(t,r)$, which is determined by the geodesic equation:
\beq
k^{\mu}\nabla_{\mu}k^{\nu}=0\;\; \Rightarrow \;\; 2F\dot{a}+a\dot{F}+F'=0,
\eeq
where $'\equiv\partial_{r}$. The geodesic expansion is
\beq
\Theta:=\nabla_{\mu}k^{\mu}=3\dot{a}F+a\dot{F}+F'+\fr{F}{r}=F\left(\dot{a}+\fr{1}{r}\right),
\eeq
where the last equality follows from the geodesic equation. A trapped surface will occur when
\beq
\dot{a}\leq-\fr{1}{r}. \label{tr}
\eeq
Clearly, for an expanding universe ($\dot{a}>0$) there can be no trapped surfaces, but for an imploding one ($\dot{a}<0$) such surfaces can form. An imploding universe is characterized by
\beq
a(t)=\sqrt{c_{1}t+c_{2}}, \;\;\;\; \mbox{with} \;\;\; c_{1}<0 \;\; \mbox{and} \;\; c_{2}>0,
\eeq
where $t\in(-\infty,-c_{2}/c_{1}]$. The condition for a trapped surface is then
\beq
r\geq2\fr{a(t)}{|c_{1}|},
\eeq
that is, for a given spacelike slice $t=t_{*}$ the region $r\geq r_{*}\equiv2\sqrt{c_{1}t_{*}+c_{2}}/|c_{1}|$ is trapped. The trapped surface $r=r_{*}$ is thence the {\em inner} boundary of a region that extends all the way to asymptotic spatial infinity (thus without compact outer boundary). It follows that neither $r=r_{*}$, nor any $r>r_{*}$ can be the outer boundary of a closed trapped region, i.e., there are no $S^{1}$ apparent horizons in the spacetime.

That the trapped region is inner trapped without a closed outer boundary (as opposed to an apparent horizon, which is the outer boundary of a compact trapped region) can also be made clear via a Penrose diagram. To see this, introduce a conformal time coordinate:
\beq
\eta:=\int \fr{dt}{a(t)}=2\fr{a(t)}{c_{1}},
\eeq
where we set $\eta(0)=2\sqrt{c_{2}}/c_{1}$ without loss of generality. The metric reads then
\beq
ds^{2}=\fr{c_{1}^{2}}{4}\eta^{2}(-d\eta^{2}+dr^{2})+\left(\fr{c_{1}}{2}\eta r\right)^{2}d\phi^{2}
\eeq
Now, introduce null coordinates:
\beq
u=\fr{\eta-r}{2},\;\;\;\; v=\fr{\eta+r}{2},
\eeq
so that the metric becomes
\beq
ds^{2}=-c_{1}^{2}(u+v)^{2}dudv+\fr{c_{1}^{2}}{4}(v^{2}-u^{2})^{2}d\phi^{2}
\eeq

In $(u,v,\phi)$ coordinates, a future-oriented, outgoing radial null geodesic is given by
\beq
k^{\mu}=(0,f(u,v),0),
\eeq
where $f>0$ is given by the geodesic equation:
\beq
(u+v)f_{,v}+2f=0.
\eeq
The geodesic expansion is
\beq
\Theta=\fr{vf}{v^{2}-u^{2}}.
\eeq
So the condition for trapped surfaces is
\beq
\fr{v}{v^{2}-u^{2}}\leq0\;\; \Rightarrow \;\; v\geq0 \;\; \mbox{and} \;\; u<v<-u,
\eeq
where we used the conditions $r>0$ and $\eta<0$. The surface $v=0$ is a trapped surface, but {\em it is not an apparent horizon} since the trapped region $v>0$ is {\em outside} the $v=0$ null surface. The $v>0$ trapped region has the $v=0$ as its {\em inner boundary}, and extends all the way to asymptotic spacelike infinity on any given spacelike slice $t=\mbox{const.}$

To construct the Penrose diagram, introduce new coordinates:
\beq
\psi=\tan^{-1}u+\tan^{-1}v, \;\;\;\; \xi=\tan^{-1}v-\tan^{-1}u.
\eeq
The center of symmetry and the curvature singularity are given by, respectively:
\bqa
r&=&0 \;\; \Rightarrow \;\; u=v \;\; \Rightarrow \xi=0, \\
a&=&0 \;\; \Rightarrow \;\; u=-v \;\; \Rightarrow \psi=0.
\eqa
In the diagram, the $\psi$ and $\xi$ axes (not shown) are vertical and horizontal, respectively. The center $r=0$ becomes singular at $u=v=0$, which corresponds to $\psi=\xi=0$; this is the left upper corner of the Penrose diagram. Collapse starts from asymptotic past timelike infinity and ends up in a spacelike singularity at $t=-c_{2}/c_{1}$.

\begin{figure}
\begin{center}
\epsfysize=20pc
\epsfxsize=30.212pc
\epsffile{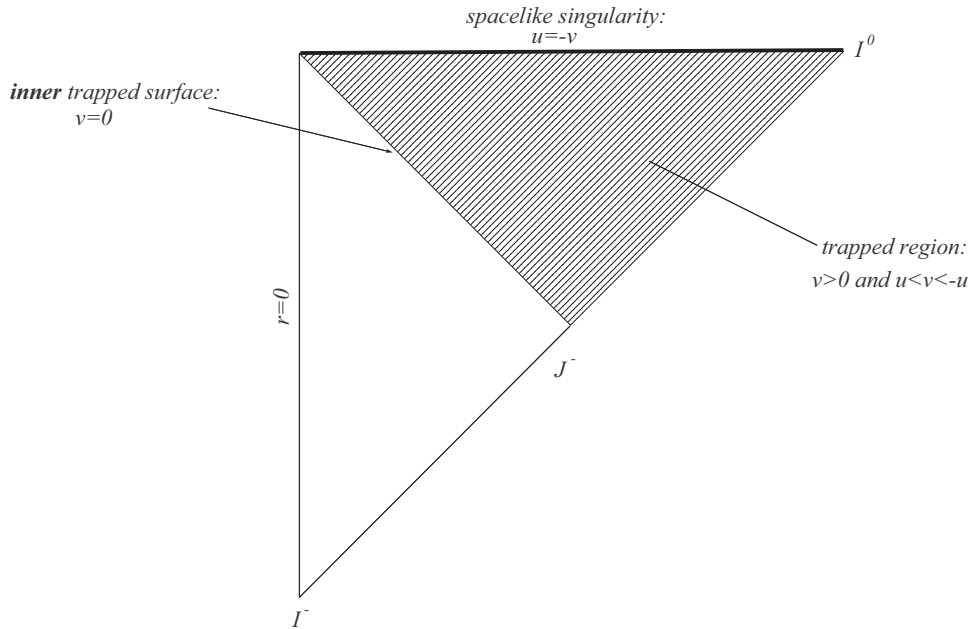}
\end{center}
\caption{Penrose diagram for collapsing flat FRW universe with $p=\rho$. The $\phi$ coordinate has been suppressed, so that each point is to be multiplied by $S^{1}$. The trapped region has an inner boundary at $v=0$, but no outer boundary, extending all the way to asymptotic spacelike infinity. This spacetime does not contain apparent horizons. \label{fig1}}
\end{figure}

Consider now Ida's theorem, which essentially states that {\em there are no apparent horizons in $(2+1)$ gravity coupled to matter which obeys the dominant energy condition}~\cite{ida00}. It is trivial to check that the stress tensor (\ref{stress}) obeys the dominant energy condition; thus, if one were to adopt W's definition of apparent horizon, the $v=0$ surface would be an apparent horizon, and this simple model would disprove Ida's theorem~\footnote{I am grateful to D. Ida for discussions regarding this point.}. This is, of course, {\em not} the case, because the ``apparent horizon'' in Ida's theorem---as in all other results in classical general relativity---is an {\em outer} boundary, not an inner one.

\subsection*{Further comments}

$\bullet$ In the second paragraph of the Introduction in \cite{wang03}, W writes ``Basing on two concrete examples, on the other hand, the author also conjectured that {\em realistic matter is required to prevent singularity formation.} In the conjecture, {\em realistic} was understood as that the weak energy condition (WEC) holds and at least one of the principal pressures is non-vanishing. Although it is not our purpose to find counter-examples to this conjecture, we would like to point out that such examples already exist, in which a singularity can be formed on the symmetry axis from the gravitational collapse of a cylindrically symmetric thin shells that consist of counter-rotating dust particles. In these models, WEC holds and the principal tangential pressure is different from zero, so all the conditions required in the conjecture are satisfied.''

Contrary to what W claims, no such examples exist. This is a simple matter of {\em necessary} versus {\em sufficient} conditions. The conjecture states that realistic matter (in the sense described above) is {\em required} to prevent singularity formation. That is, realistic matter is a {\em necessary condition}, not a sufficient one, in the statement of the conjecture. It should therefore be quite clear that examples with such realistic matter which develop singularities do {\em not} constitute counter-examples to the conjecture. 

\vspace{0.2cm}

$\bullet$ In the third paragraph of the Conclusions in \cite{wang03}, W writes ``At this point some comments on the hoop conjecture is in order. One may particularly ask: Is the example presented here in conflict with the conjecture? because according to it horizons cannot be formed from gravitational collapse with cylindrical symmetry. To answer this question, the first thing one may need to do is to show that the trapped surfaces presented in the above example can be formed from gravitational collapse.''

It seems to me that, contrary to what W proposes, there is no need to ``show that the trapped surfaces presented in the above example can be formed from gravitational collapse.'' This is because such surfaces are not apparent horizons in the standard sense, as previously discussed in this note. Even if trapped surfaces (of the kind described by W) were to develop during gravitational collapse, they would not represent {\em confinement} of mass; quite the opposite, one needs to go to larger radii to encounter a trapped surface. This is the case of the $(2+1)$ model previously discussed here, and it also happens in several $(3+1)$ cosmological Friedman models and spherical homogeneous dust collapse~\footnote{One must note that the homogeneous dust metric (the well-known Oppenheimer-Snyder metric) is only valid in the matter filled region, which is to be matched to a vacuum Schwarzschild exterior (such matching is uniquely defined, given any constant density value for the interior dust region). Because the exterior is Schwarzschild, the trapped region will not extend all the way to spacelike infinity on any given Cauchy slice (as it would if the spacetime was not asymptotically flat); instead, it will be outer bounded by the event horizon.}: outer shells become trapped before inner shells because what causes the ``trapping'' is the total mass inside a given radius, and (because of the density homogeneity or special equation of state) one needs larger radii to achieve such trapping. In very crude terms, collapse is just ``too slow'' to be able to concentrate enough mass inside the inner shells.

The spirit of the hoop conjecture is the notion that mass (be it matter and/or gravitational energy, as in the case of gravitational waves) must be {\em sufficiently compacted along all of the three spacelike directions} in order for a ``horizon'' to form. Thorne's original statement of the conjecture~\cite{thorne72} is deliberately vague in terms of the definitions of mass, horizon, and circumference. Despite such ambiguity, however, there does not appear to be any credible counter-example to date. Clearly, examples of collapsing spacetimes with trapped regions which are inner but not outer bounded, do not constitute counter-examples to the hoop conjecture.

\end{document}